# Separation of bacterial spores from flowing water in macro-scale cavities by ultrasonic standing waves

B. Lipkens, J. Dionne
Western New England College
Department of Mechanical Engineering
1215 Wilbraham Road, Box S-5024
Springfield, Massachusetts 01119
blipkins@wnec.edu

M. Costolo, A. Stevens, Edward Rietman
Physical Sciences Inc.
20 New England Business Center
Andover, Massachusetts 01810
costolo@psicorp.com, stevens@psicorp.com, erietman@gmail.com

**Abstract**

The separation of micron-sized bacterial spores (*Bacillus cereus*) from a steady flow of water through the use of ultrasonic standing waves is demonstrated. An ultrasonic resonator with cross-section of 0.0254 m x 0.0254 m has been designed with a flow inlet and outlet for a water stream that ensures laminar flow conditions into and out of the resonator section of the flow tube. A 0.01905-m diameter PZT-4, nominal 2-MHz transducer is used to generate ultrasonic standing waves in the resonator. The acoustic resonator is 0.0356 m from transducer face to the opposite reflector wall with the acoustic field in a direction orthogonal to the water flow direction. At fixed frequency excitation, spores are concentrated at the stable locations of the acoustic radiation force and trapped in the resonator region. The effect of the transducer voltage and frequency on the efficiency of spore capture in the resonator has been investigated. Successful separation of *B. cereus* spores from water with typical volume flow rates of 40-250 ml/min has been achieved with 15% efficiency in a single pass at 40 ml/min.

## I.     Introduction

We previously reported[1] the use of ultrasonic standing waves for the separation of 6-micron diameter polystyrene beads from a flowing water sample as the water passed through a 0.15-m long resonator cavity placed symmetrically across a 0.0254-m x 0.0254-m flow channel. That system relied on a frequency sweep method to translate the particles into the arm of the resonator cavity that is opposite the acoustic transducer. The collection efficiency of that resonator and flow channel design was limited significantly by the acoustic streaming of the water as a result of the frequency sweep. We report here an alternate system design in which the applied ultrasonic frequency is kept fixed, the resonator cavity is much shorter (0.0356m from transducer face to the opposing reflective wall), and the separation of micron-sized bacterial spores is achieved simply by their capture in the stable locations of the acoustic radiation field.

There is a large body of work on the use of the acoustic radiation force for particle concentration and separation. A recent review paper by Marston and Thiessen[2] presents an overview of the fundamental principles and applications of the acoustic radiation force. The theoretical framework has been developed by King,[3] Yosioka and Kawasima,[4] Gor'kov,[5] and Tolt and Feke.[6,7] They developed a separation process based on the acoustic radiation force in a stationary ultrasonic standing wave field. Tolt and Feke[6] used a frequency sweeping method to translate the concentrated particles across the resonator. The frequency sweep is over a range of $2f_0$, where $f_0$ is the fundamental resonance frequency of a resonator with length L, i.e., $f_0=2c_f/L$, where $c_f$ is the speed of sound of the fluid in the resonator. Hill et al.[8,9] and Townsend et al.[10] developed a model for the calculation of particle paths for suspended particles in a fluid. The particles are exposed to three forces, acoustic radiation force, fluid drag force, and buoyancy force. An electro-acoustic model is used to calculate the acoustic field in the resonator. A separate model is used to calculate the flow velocity vectors in the resonator. A third model considers the particle-acoustic field interaction. A second example of using frequency sweeps to translate particles across an ultrasonic resonator is presented in the work by Handl et al.[11] They used a four step frequency sweep that alternated between the frequencies of four consecutive acoustic resonance frequencies of the cavity. They showed that under these conditions particles are translated across the resonator. The measured particle trajectories compared favorably with those predicted by a numerical model.

## II.    Acoustic Radiation Force

For design calculations a particle trajectory model has been developed.[1] In this model the acoustic field generated by a piezoelectric transducer is predicted. The acoustic radiation force is then calculated according to the formulation of Gor'kov.[5] The primary acoustic radiation force $F_A$ is defined as a function of a field potential U

$$F_A = -\nabla(U), \qquad (1)$$

where the field potential U is defined as

$$U = V_0 \left[ \frac{\langle p^2(x,y,t) \rangle}{2\rho_f c_f^2} f_1 - \frac{3\rho_f \langle u^2(x,y,t) \rangle}{4} f_2 \right], \tag{2}$$

and $f_1$ and $f_2$ are the monopole and dipole contributions defined by

$$f_1 = 1 - \frac{1}{\Lambda \sigma^2},$$
$$f_2 = \frac{2(\Lambda - 1)}{2\Lambda + 1} \tag{3}$$

where $p(x,y,t)$ is the acoustic pressure, $u(x,y,t)$ is the fluid particle velocity, $\Lambda$ is the ratio of particle density ($\rho_p$) to fluid density ($\rho_f$), $\sigma$ is the square of the ratio of particle sound speed ($c_p$) to fluid sound speed ($c_f$), and $V_o$ is the volume of the particle. For a one-dimensional harmonic acoustic wave, the expression for the acoustic radiation force is:

$$F_{Ax} = V_0 \left[ \frac{3(\rho_p - \rho_f)}{2\rho_p + \rho_f} \frac{\rho_f}{2} U_M \frac{dU_M}{dx} \right] -$$
$$V_0 \left[ \left(1 - \frac{\rho_f c_f^2}{\rho_p c_p^2}\right) \frac{1}{2\rho_f c_f^2} P_M \frac{dP_M}{dx} \right], \tag{4}$$

where $U_M$ and $P_M$ are the magnitudes of acoustic velocity and pressure. The calculation of the acoustic radiation force is implemented through a numerical calculation of the spatial derivative of velocity and pressure. The model is described in detail in Lipkens et al.[12] Understanding the numerical model is important for this work in that it illustrates the difficulty of capturing bacterial spores: they have extremely small volumes ($V_0$ about 3 x $10^{-19}$ m$^3$), a density close to that of water ($\rho_p$ estimated at 1.07 g/cm$^3$)[13] which results in a small acoustic radiation force on the particle.

### III. Design and Experimental Setup

#### A. Acoustocollector design

The acoustocollector, consisting of a flow cell and resonator, is shown in Fig. (1). The flow inlet and outlet channels are each 0.0254 m x 0.0254 m in cross-section and are attached to the resonator. The length of the inlet and outlet is designed to be at least the entrance length related to the Reynolds number associated with typical flow rates of 150 ml/min, thereby ensuring laminar flow conditions in the acoustoresonator; lengths of the inlet and outlet channels were anywhere from 0.10 to 0.20 m. The inlet and outlet were manufactured in an Objet rapid prototype machine.

The ultrasonic resonator section is machined of aluminum; the front face is made of glass to allow visual observation. The resonator is a 0.0508-m long section inserted between the inlet and outlet sections, and has a square cross-section of 0.0254 m x 0.0254 m to match the inlet and outlet cross sections. The transducer is in direct contact with the water and is mounted in the top face of the resonator, so that the acoustic field direction is orthogonal to the flow direction. A collection pocket, 0.0254 m x 0.0254 m and 0.0102 m deep is in the resonator face that opposes the transducer; the resonator cavity therefore has a face-to-face distance of 0.0356 m. To enable gravitational settling of the spores into the collection pocket, the system is mounted so the flow is horizontal, as shown in Fig. (1).

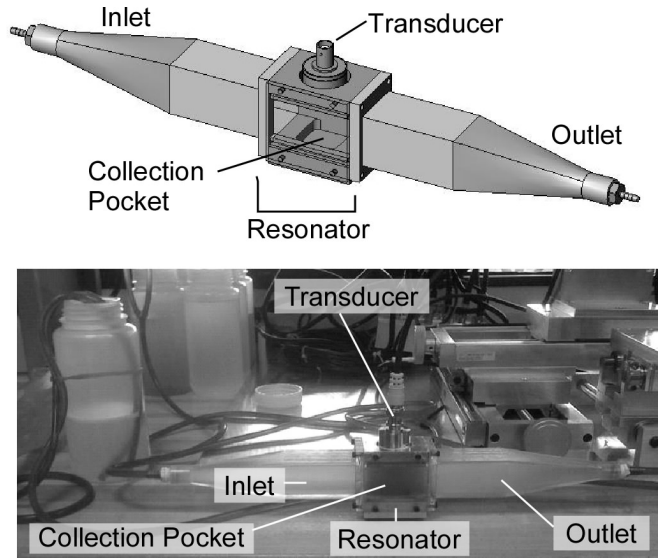

**Fig. 1. a) Isometric view and b) photograph of the acoustocollector, showing the inlet and outlet flow channels, resonator cavity, transducer, and collection pocket.** The flow enters from the left. The acoustic field is in the vertical direction, with the transducer mounted on the top face of the resonator. The action of the ultrasonic radiation force is so as to trap the passing particles in the field, as well as cause agglomeration. The spores can be harvested from the collector pocket.

The transducer was manufactured by UTX Inc., and is PZT-4 with a nominal resonance frequency of 2 MHz; its face has a diameter of 0.75-in (0.01905m). Similar 2-MHz transducers were also purchased from The Ultran Group. The S11 curve (ratio of transmitted to reflective power as a function of the excitation frequency) for each of the various transducers was determined using an Agilent E8356A vector network analyzer with the transducer face immersed in water in the cavity of Fig. (1). A Tektronix function generator was used to generate the drive signal which was amplified by an Amplifier Research 100-W amplifier with a bandwidth from 10 kHz to 250 MHz. Typical voltages applied across the PZT-4 were in the range of 20 to 40 Vpp. A dual-head Cole-Palmer L/S Economy Variable-Speed Digital-Drive Peristaltic Pump is used to pump the water through the resonator. Flow rates used in the experiments varied from 40 to 250 ml/min.

### B. Diagnostic apparatus

Two techniques were used to qualify the acoustocollector. The first is simply qualitative, and consists of a ProScope USB microscope with a 10x magnifying lens that can be placed so as to image the captured spores in the resonator, and store the video images on a computer.

Quantitative determination of the relative concentration of bacterial spores was achieved by using in-line optical scattering, shown in Fig. (2).

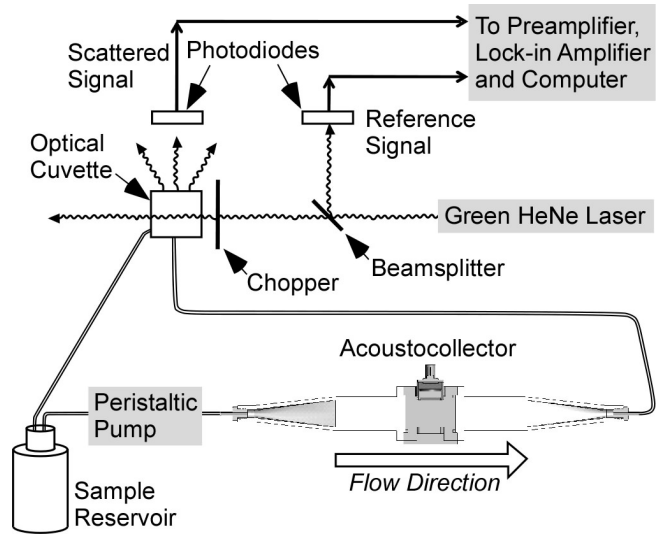

**Fig. 2. Schematic of the optical-scattering apparatus used for in-line determination of relative spore concentrations.**

A green HeNe laser beam goes through a beamsplitter with the split beam picked up by a photodiode and used as a laser power reference to factor in any fluctuations in the laser intensity. The forward beam is chopped and then is incident on a cuvette which has the outlet water stream from the acoustocollector suspension flowing through it. The light scattered at 90-degree scattering signal is picked up by a photodiode; the signal goes to a preamplifier and then to a lock-in amplifier. The resulting light intensity data from reference and signal photodiodes are captured using an A/D board and custom LabVIEW software.

### C. Preparation of spore samples

*B. cereus* spores were prepared by growing the bacteria (ATCC 14579) in nutrient medium at 30°C for three days, followed by collection of the spores by centrifugation. The spores were thoroughly washed with sterile deionized water and stored at 4°C for several weeks in sterile DI water. Occasional centrifugation of the sample and re-suspension in water during this time removed any residual cellular matter. Spores were stained by adding a small amount of Malachite Green to the sample in water, heating for about 10 min. to 80-85°C, and then repeatedly washing the stained spores to remove excess dye.

A photomicrograph of a sample of the spores is shown in Fig. (3). The *B. cereus* spores are cylindrical, about 1.5 μm in length and 0.5 μm in diameter (volume $V_0 = 2.9 \times 10^{-19}$ m$^3$); their density is estimated to be 1.07 g/cm$^3$;[13] their compressibility has not been measured.

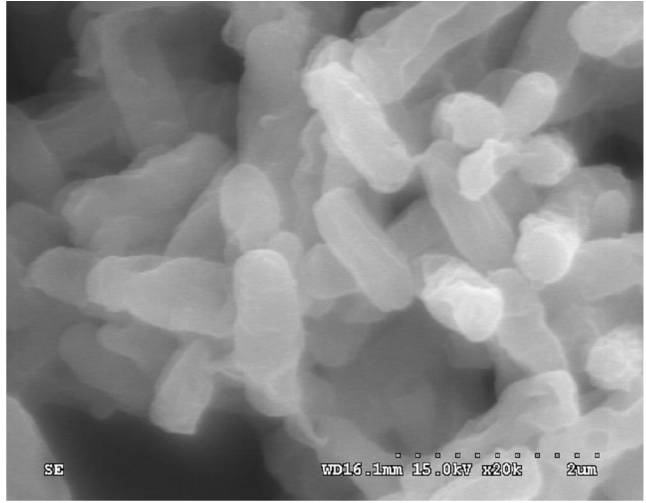

**Fig. 3. Scanning electron micrograph of a preparation of *B. cereus* spores; the spores are about 1.5 μm in length and 0.5 μm in diameter.**

Dilute samples of spores in water were prepared using 0.05% Tween 20, a surfactant that prevents the spores from agglomerating in the stock solutions. Spore concentrations for these experiments were about 10$^7$ per mL.

## IV. Results

### A. Collection of *B. cereus* in the acoustocollector

A 500-ml suspension of *B. cereus* spores in water (with 0.05% Tween) was pumped at a relatively slow flow rate of 40 ml/min through the acoustocollector system shown in Figs. (1) and (2). A column of captured spores forms in the resonator when power is applied to the transducer at a constant frequency which is a cavity resonance; a photomicrograph of a column of captured spores is shown in Fig. (4). The column formation occurs with an ~2.2 MHz transducer frequency and an applied voltage of 20Vpp across the transducer. As the system comes to thermal equilibrium, the applied frequency must be adjusted (increased) to compensate for the change in the speed of sound of the fluid with increasing temperature. In Fig. (4) the spores are stacked in planes parallel to the transducer face; the planes correspond to the stable locations of the zeroes of the acoustic radiation force. Additionally, there exists a radial component of the acoustic radiation force that holds the spores in these planes against the fluid flow. A batch process can be implemented where spores are trapped and held until some volume of water has been processed, and the flow of water and power to the transducer turned off so that the agglomerated spores fall under the force of gravity to the collector pocket. Alternatively, a slow frequency sweep could be used to drive the particles into the collector pocket.

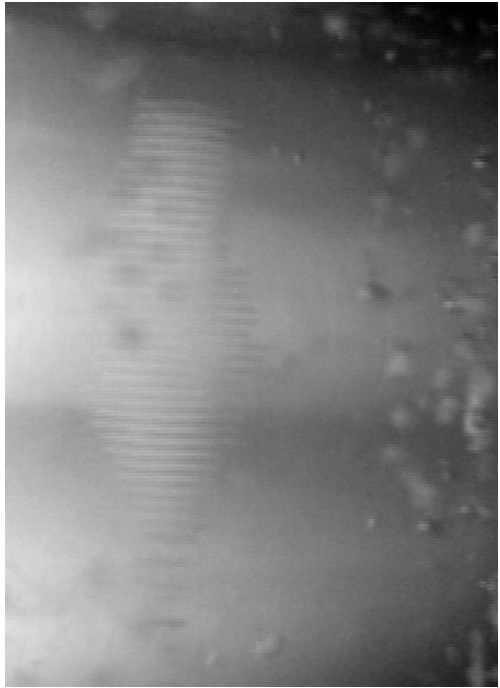

**Fig. 4. Microphotograph of a column of captured *B. cereus* spores stacked in planes about 700 microns apart in the resonator.** The spores have been dyed with Malachite Green to make them more clearly visible. The orientation of the acoustocollector is as shown in Fig. (1); the transducer is at the top, just out of the field of view and facing downward, and the water flow is moving to the right at 40 ml/min.

Fig. (5) shows the optical scattering signal from the outlet stream, normalized to the initial scattered intensity (proportional to the starting spore concentration); the x-axis units are in terms of bottle turnovers, since each bottle consists of a 500-ml sample volume; the outlet from the collector is fed back into the sample bottle. Fig. (5) indicates that in the initial pass through the system, ~15% of the spores were trapped. Note, however, the reduction in slope at approximately 0.5 bottle turnovers; this decrease in capture efficiency was accompanied by a spontaneous increase in the voltage measured across the transducer from $20V_{pp}$ to $30V_{pp}$ (constant driver voltage and amplifier gain, i.e., applied power), which occurred even with adjustment of the applied frequency.

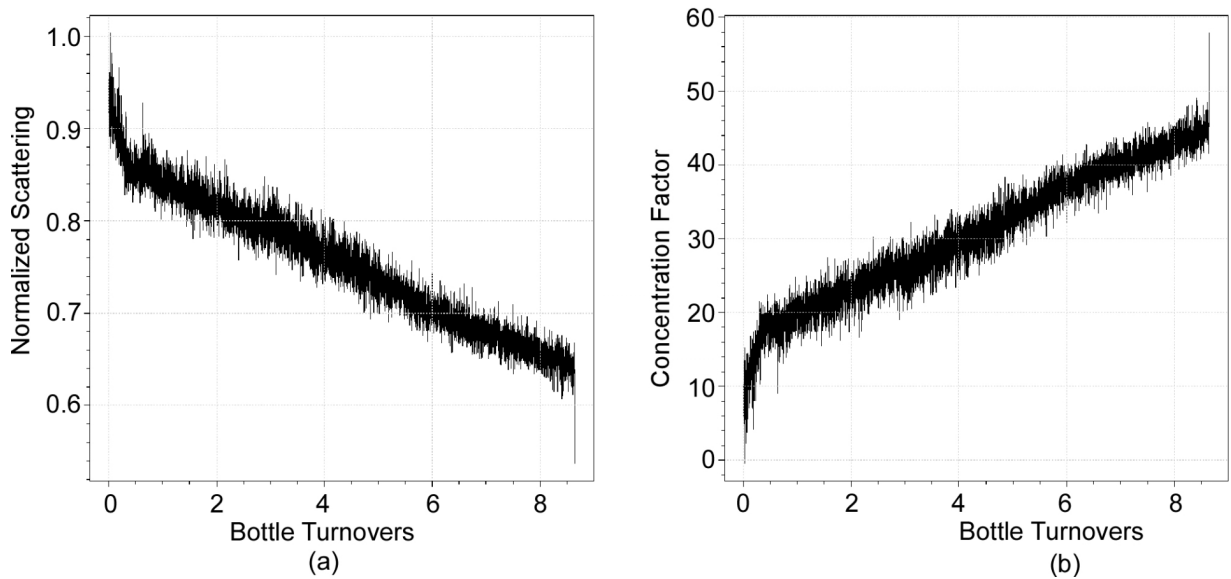

**Fig. 5. a) The scattering intensity in the outlet stream as a function of bottle turnovers, with the intensity at $t=0$ set equal to one for normalization. b) The concentration factor $C$ as given by Eq. (5), assuming a 4-ml collection volume for the captured spores.** Data were taken with the transducer frequency set to 2.2 MHz (with slight re-tuning of the frequency as the temperature of the sample increased), a 20 $V_{PP}$ initial applied voltage (which shifted to about 30 $V_{PP}$ around 0.5 a bottle turnover), a flow rate set to 40 ml/min, and an initial spore concentration of about $10^7$ per ml.

The concentration factor $C$ is calculated from the total fraction of the spores captured ($n$), the total volume of processed fluid V and the volume of fluid v in the collection pocket.

$$C = (1-n)(V/v) \qquad (5)$$

Fig. (5)b shows the concentration factor $C$ as a function of bottle turnovers for a 500-ml sample volume ($V$) and a 4-ml collection volume ($v$), and shows that a concentration increase of about a factor of 50 was achieved after 8 passes through the concentrator. After flowing the spore suspension through the acoustocollector 8 times, the flow was stopped and then the power to the transducer turned off. The captured spores readily fell into the collection pocket at the bottom of the resonator cavity, where they were removed using a pipet.

In order to investigate the spore capture as a function of the applied transducer frequency and the resonator modes, the transducer frequency was scanned from 2.0 to 2.3 MHz in 1 kHz steps for a fixed function generator output and fixed amplifier gain (i.e., constant applied power). During the scan the voltage across the transducer was measured; simultaneously the spore capture was monitored with the digital microscope videocamera so that conditions producing a column could be noted. Fig. (6) shows the results of the frequency scan for a transducer that was characterized by a S11 minimum at 2.211 MHz (in water, external to the resonator).

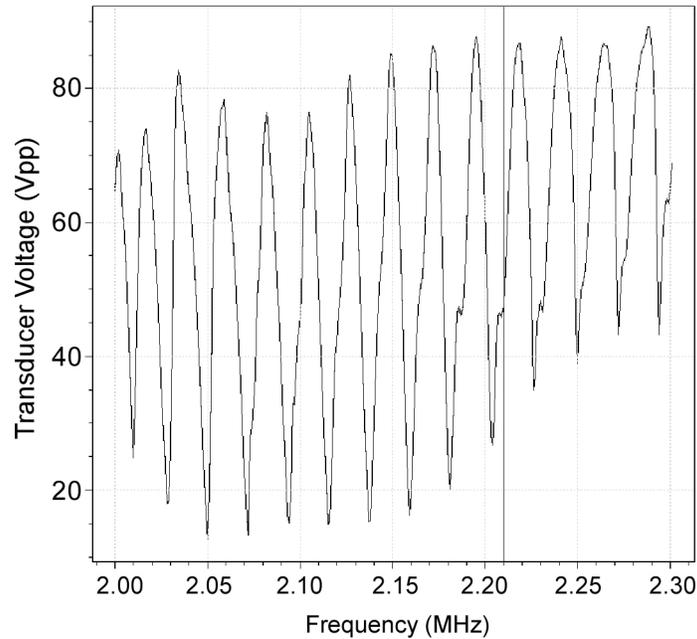

**Fig. 6. Measured transducer voltage vs. drive frequency for a fixed amplifier gain (total system power) for the transducer/resonator in the acoustocollector.** The S11 minimum for this particular transducer in water (with no resonant cavity) was measured in a separate experiment and found to be at 2.211 MHz, indicated by the vertical line.

Spore capture tends to be initiated at frequencies corresponding to local minima in the aluminum transducer voltage as a function of applied frequency (i.e., frequencies in resonance with the aluminum resonator and that therefore result in standing waves), and at frequencies closest to the S11 minimum (i.e., where the most power is delivered from the amplifier to the transducer itself). The column of captured spores remains stable if the frequency is then tuned to a region that produces higher voltages across the transducer, but it does not begin formation as readily at the frequencies corresponding to local maxima in the graph of voltage vs. frequency. Column formation starts a few minima to the low frequency side of the S11 frequency (2.211 MHz) and continues through the S11 minimum. The frequency ranges that include a secondary minimum on the high-frequency side tend to cause intense bubble formation, presumably due to very high acoustic pressures that result from the applied frequency being near the transducer resonance. Once bubbles begin to form in the column, spore collection stops. We attribute this to the presence of the air bubbles effectively damping the acoustic waves, due to the large compressibility difference of air and water.

Effective spore capture takes place when the highest acoustic pressure is obtained, but without creating bubbles. For example, for the transducer/resonator system of Fig. (6), this corresponds to a frequency of ~2.16 MHz, three minima to the lower frequency side of the S11 resonance. The voltage vs. frequency curve of Fig. (6) shifts to higher frequencies with increasing temperature, so the frequency corresponding to the voltage minimum must be found experimentally each time an experiment is run.

## V. Conclusions

Successful separation of bacterial spores from a 40-250 ml/min water stream using a fixed acoustic frequency is demonstrated in an acoustocollector with a 0.0508-m long aluminum acoustic resonator section inserted between inlet and outlet sections. A 0.01905-m diameter PZT-4, nominal 2-MHz transducer is used to generate ultrasonic standing waves in the resonator and capture the spores in the resulting acoustic pressure fields; the radial field holds the spores against the water flow.

*B. cereus* spores can be captured with 15% capture efficiency in a single pass through an acoustic resonator at 40ml/min flow rate. The resulting agglomeration of spores can be collected by stopping the water flow and the transducer, allowing gravitational settling of the spores, and removing the spore sample from a collection pocket at the bottom of the resonator.

The acoustocollector is ideally suited for large-volume sampling of water supplies for concentration of spores and transfer of the resulting concentrated sample to apparatus—e.g., Raman or InfraRed spectrometers—for further identification and analysis.

## Acknowledgements

This work was funded by the U.S. Army under STTR contract number W911SR-06-C-0040. We thank Dr. James Jensen for technical discussions, I. Konen for assistance assembling the laser scattering apparatus, J. Towle for assistance measuring the transducer S11 curve, A. Ferrante, D. Vu, and L. Harrison for advice and assistance in spore preparation, and B. Szczur and A. Trask for testing assistance at Western New England College.
## References

1. B. Lipkens, J. Dionne, A. Trask, B. Szczur, A. Stevens, and E. Rietman, "Separation of micron-sized particles in macro-scale cavities by ultrasonic standing waves," International Congress on Ultrasonics, Santiago de Chile, January 2009, in: Physics Procedia, 00,0000-0000, 2009.
2. P. L. Marston and D. B. Thiessen, "Manipulation of fluid objects with acoustic radiation pressure," Ann. N. Y. Acad. Sci. **1027**, 414-434 (2004).
3. L. V. King, "On the acoustic radiation pressure on spheres," Proc. R. Soc. London, Ser. A **147**, 212-240 (1934).
4. K. Yosioka and Y. Kawasima, "Acoustic radiation pressure on a compressible sphere," Acustica **5**, 167-173 (1955).
5. L. P. Gor'kov, "On the forces acting on a small particle in an acoustical field in an ideal fluid," Sov. Phys. Dokl. **6**, 773-775 (1962).
6. T. L. Tolt and D. L. Feke, "Separation devices based on forced coincidence response of fluid-filled pipes," J. Acoust. Soc. Am. **91**(6), 3152-3156 (1992).
7. T. L. Tolt and D. L. Feke, "Separation of dispersed phases from liquids in acoustically driven chambers," Chem. Eng. Sci. **48**(20), p. 527 (1993).
8. M. Hill and R. J. K. Wood, "Modelling in the design of a flow-through ultrasonic separator," Ultrasonics **38**, 662-665 (2000).

9. M. Hill, Y. Shen, and J. J. Hawkes, "Modelling of layered resonators for ultrasonic separation," Ultrasonics **40**, 385-392 (2002).
10. R. J. Townsend, M. Hill, N. R. Harris, and N. M. White, "Modelling of particle paths passing through an ultrasonic standing wave," Ultrasonics **42**, 319-324 (2004).
11. B. Handl, M. Groschl, F. Trampler, E. Benes, S. M. Woodside, and J. M. Piret, "Particle trajectories in a drifting resonance field separation device," Proc. of the 16th Int. Congress on Acoustics and 135th Meeting of the Acoust. Soc. Am., ISBN 1-56396-817-7, vol. III, pp. 1957-1958 (1998).
12. B. Lipkens, M. Costolo, and E. Rietman, "The effect of frequency sweeping and fluid flow on particle trajectories in ultrasonic standing waves," IEEE Sensors Journal **8**(6), pp. 667-677 (2008).
13. Y. Huang, X.-B. Wang, F. F. Becker, and P. R. C. Gascoyne, "Introducing Dielectrophoresis as a New Force Field for Field-Flow Fractionation," Biophys. J. **73**(2), pp. 1118-1129 (1997).